\documentclass[12pt]{article}
\usepackage{graphicx,psfrag,epsf}
\usepackage{hyperref}
\usepackage{enumerate}
\usepackage{natbib}
\usepackage{url} 
\usepackage{hhline,array,amsmath,graphicx}
\usepackage{pifont,latexsym,ifthen,rotating,calc}
\usepackage{siunitx}
\usepackage{etoolbox}
\usepackage{mathtools}
\usepackage{amsmath}
\usepackage{amssymb}
\usepackage{amsthm}
\usepackage{bbm}
\usepackage{pdflscape}
\usepackage{rotating}
\usepackage{algorithm}
\usepackage{algpseudocode}
\usepackage{bm}
\usepackage{enumitem}
\usepackage[mathscr]{euscript}
\usepackage{multirow}
\usepackage{xcolor}
\usepackage{bm}
\usepackage{amsmath}
\usepackage{amsfonts}
\usepackage{multirow}
\newcommand{\widesim}[2][1.5]{
  \mathrel{\overset{#2}{\scalebox{#1}[1]{$\sim$}}}
}

\newcommand{\blind}{0}

\begin{document}

\def\spacingset#1{\renewcommand{\baselinestretch}%
{#1}\small\normalsize} \spacingset{1}

\date{}


\if0\blind
{
  \title{\bf Fiducial Inference for Random-Effects Calibration Models: Advancing Reliable Quantification in Environmental Analytical Chemistry}
  \author{Soumya Sahu\\
    Division of Epidemiology and Biostatistics, University of Illinois Chicago\\
    Thomas Mathew\\
    Department of Mathematics and Statistics, University of Maryland Baltimore County\\
    Robert Gibbons\\
    Department of Public Health Sciences, The University of Chicago\\
    Dulal K. Bhaumik\\
    Division of Epidemiology and Biostatistics, University of Illinois Chicago\\
    Department of Psychiatry, University of Illinois Chicago}
  \maketitle
} \fi

\if1\blind
{
  \bigskip
  \bigskip
  \bigskip
    \title{\LARGE\bf Fiducial Inference for Random-Effects Calibration Models: Advancing Reliable Quantification in Environmental Analytical Chemistry}
  \maketitle
} \fi

\bigskip

\begin{abstract}
    
This article addresses calibration challenges in analytical chemistry by employing a random-effects calibration curve model and its generalizations to capture variability in analyte concentrations. The model is motivated by specific issues in analytical chemistry, where measurement errors remain constant at low concentrations but increase proportionally as concentrations rise. To account for this, the model permits the parameters of the calibration curve, which relate instrument responses to true concentrations, to vary across different laboratories, thereby reflecting the potential variability in measurement processes. The calibration curve that accurately captures the heteroscedastic nature of the data results in more reliable estimates across diverse laboratory conditions. Noting that traditional large-sample interval estimation methods are inadequate for small samples, an  alternative approach, namely the fiducial approach, is explored in this work.  It turns out that the fiducial approach, when used to construct a confidence interval for an unknown concentration, outperforms all other available approaches in terms of maintaining the coverage probabilities. Applications considered include the determination of the presence of an analyte and the interval estimation of an unknown true analyte concentration. The proposed method is demonstrated for both simulated and real interlaboratory data, including examples involving copper and cadmium in distilled water.
\end{abstract}
\noindent%
{\it Keywords:} Interlaboratory analysis, Fiducial quantity, Hierarchical designs, Confidence width, Rocke-Lorenzato model. 
\vfill

\def\spacingset#1{\renewcommand{\baselinestretch}%
{#1}\small\normalsize} \spacingset{1}


\newpage
\spacingset{2} 
\section{Introduction}
\label{sec:intro}
A critically important problem in interlaboratory calibration is the accurate interval estimation of the unknown true concentration of an analyte in the context of quantitative analysis in analytical chemistry. The problem is especially challenging  when dealing with heterocedastic measurement errors at different concentration levels of the analyte. In practice, analytical measurements are routinely used for various purposes, often without considering the associated level of uncertainty. For instance, in environmental monitoring programs, the concentration of low-level environmental pollutants in a single sample is directly compared with regulatory standards without accounting for the uncertainty in those measurements (see \cite{gibbons1994statistical} for reviews of statistical issues in environmental monitoring applications). The uncertainty can be quantified by providing confidence bounds for the true concentration level, given measured concentrations obtained under a hierarchical structure, such as concentrations measured across different laboratories. It is common in environmental monitoring to use multiple laboratories for routine monitoring or to cross-validate an anomalous result. This nested structure, where the laboratory is considered a random factor, complicates the computation of confidence bounds for the calibration problem. 

The practical importance of the problem is two-fold. First, in a standard situation, a particular laboratory  provides an estimate of an unknown concentration, in rare cases, however, the laboratory may  compute a corresponding confidence interval for its use (see \cite{rock1995}). The laboratory could benefit by borrowing strength from other laboratories that use the same analytical procedure. Both the point estimate and confidence interval would be more realistic if they accounted for interlaboratory variation, thereby reducing the potential bias of the laboratory. Second, it is common in routine analytical practice for a sample to be ``split" into a series of subsamples sent to different laboratories (e.g., a state laboratory, a commercial laboratory, and a U.S. Environmental Protection Agency [EPA] laboratory). The methodology presented here provides an estimate of the true concentration and a corresponding confidence interval based on data from multiple laboratories, borrowing strength from all laboratories involved. 

In terms of models used for interlaboratory calibration, the Rocke and Lorenzato model has become a widely adopted model in analytical chemistry \cite{rock1995}. This model incorporates both additive and multiplicative error components, and the model effectively accommodates heteroscedasticity in measurement errors across the calibration range. Specifically, it accounts for a proportional increase in error with increasing analyte concentration, and a nearly constant error at low analyte levels. This dual-component approach has demonstrated robust fitting across a broad range of analyte concentrations, and the model has garnered significant attention in the literature, leading to various adaptations and extensions; \cite{gibbons2001weighted}, \cite{bhaumik2005confidence}, and \cite{zhao2021accurate}. 

While originally formulated for single-laboratory data, adaptations of the Rocke-Lorenzato model have extended its utility to scenarios involving data from multiple laboratories. 
\cite{gibbons2001weighted}  generalized the model  to the interlaboratory scenario by using a weighted random-effects regression model to accommodate non-constant variability over the observed concentration range. This extension introduces complexities such as heteroscedasticity across laboratories, where variances differ across labs. Calibration, a primary application of the model, involves estimating unknown analyte concentrations based on the corresponding instrument responses, such as spectroscopic measurements. Further,  it builds an interlaboratory confidence region for a true concentration, given a series of new measured concentrations from multiple laboratories. Laboratory data used for estimating an unknown true concentration may be from a single laboratory,  or from a subset of the laboratories used in estimating the calibration function parameters. For example, all qualified laboratories in a state could be used for estimating the calibration function  and associated variances. Borrowing strength from these estimates, we can obtain both point and interval estimates of a true unknown concentration. 

Previous work in this area for linear models includes studies by \cite{Osborne1991, eberhardt1994constant, johnson1996combining,  cameron2022issues}. In addition, data analysis using the Rocke-Lorenzato model and its derivatives typically employs likelihood-based methods, maximizing the integrated likelihood to estimate parameters and their asymptotic variances. \cite{bhaumik2005confidence} obtained moment-based estimators of the model parameters, studied their properties, and used them to construct confidence intervals using data from multiple laboratories. However, conventional large-sample confidence intervals derived from these methods often exhibit inadequate coverage probabilities, unless sample sizes are large. In response to this limitation, \cite{zhao2021accurate} proposed a modified likelihood ratio approach using a parametric bootstrap based on a standardized signed log-likelihood ratio test to improve confidence interval accuracy. In spite its improved performance, the coverage probabilities were still unsatisfactory for some parametric choices when multiple
laboratories are involved. We shall thus explore a fiducial approach under the Rocke-Lorenzato model and its generalizations based on measurements made by a single laboratory or split samples from multiple laboratories. Based on fairly extensive simulations, we demonstrate that the fiducial approach outperforms all other available approaches, including the modified likelihood ratio procedure due to \cite{zhao2021accurate} and bootstrap-based approaches, in terms of meeting the coverage probability requirement, and yet providing confidence intervals having comparable or better expected widths. We also show that the proposed fiducial approach not only has better accuracy,  but is also computationally much more efficient than some of the competing methods, especially the methodology developed in \cite{zhao2021accurate}. The fiducial approach  has actually seen renewed interest in recent years; \cite{hannig2016} and \cite{bhaumik2021}. Earlier, \cite{weerahandi1993}, \cite{xu2008generalized}, \cite{krishnamoorthy2003lognormal} and \cite{bhaumik2013lognormal} developed and successfully applied the fiducial approach, referring to a generalized pivotal quantity. We note that the Rocke-Lorenzato model and its derivatives have random effects that enter the model in a non-linear fashion, and we have been able to develop the fiducial approach in this non-linear scenario. We have also explored the performance of the point estimate by the fiducial mode.

The paper is organized as follows: Section 2 presents two motivating applications, followed by a non-linear mixed-effects model for analyzing inter-laboratory calibration data in Section 3. Section 4 develops confidence intervals and point estimates based on the fiducial approach. Section 5 reports simulation results to compare our fiducial solutions with those based on other approaches, including the bootstrap, Wald-type large sample method, and the modified likelihood ratio method due to \cite{zhao2021accurate}. Section 6 revisits the applications. The paper concludes with a discussion in Section 7.

\section{Motivating Examples}

Our first example is an interlaboratory study to analyze cadmium concentrations in water. The example is based on a blind study carried out by the Ford Motor Company, in collaboration with Michigan State Drinking Water Certification laboratories. The study used blind samples prepared by an independent source, randomized and submitted weekly over five weeks. Cadmium was analyzed using inductively coupled plasma atomic emissions spectroscopy (EPA method 200.7). The dataset included five replicates at concentrations of 0, 20, and 100 $\mu g/L$ across five laboratories.

Our second example is based on an interlaboratory study, also conducted by the Ford Motor Company, in order to analyze copper concentration in water  (John Phillips of the Ford Motor Company, personal communication). These data were generated as part of a blind interlaboratory study among the laboratories that hold Michigan State Drinking Water Certifications for the parameters tested. Samples were prepared by an independent source, randomized, and submitted on a weekly basis over a five-week period. Copper was analyzed by Inductively Coupled Plasma Atomic Emissions Spectroscopy (ICP/AES) using Environmental Protection Agency method 200.7. The dataset consisted of five replicates at each of five concentrations ( 0 , $2,10,50$, and $200 \mu \mathrm{g} / \mathrm{L}$ ) in each of the seven laboratories.

The goals of the above projects were (a) to estimate the calibration curve for each laboratory, (b) estimate the true unknown analyte concentration, and construct corresponding  100$(1 - \alpha)$\%  confidence intervals based on data from multiple laboratories, (c) illustrate the use of the confidence interval in environmental monitoring applications where measurements are routinely compared to regulatory thresholds (RT) without incorporating uncertainty in their estimated concentration. 

The datasets are presented as Tables F.5 and F.6 in the Section 6 of the supplementary material.

\section{Choice of Model}
  Increasing measurement variation with increasing analyte concentration is commonly observed in analytical data. Denoting  the true concentration of an analyte by  $x$, the traditional simple linear calibration model specified as $y=\alpha+\beta x+e$, with the standard normality assumption on the errors, is inappropriate, as it fails to explain increasing measurement variation with increasing analyte concentration.  An alternative model can be a log-linear model-for example, $y=x e^{\eta}$, where $\eta$ is a normal variable with mean 0 and standard deviation $\sigma_{\eta}$. However, the main problem with this model is that it fails to explain near-constant measurement variation  for a low true concentration level $x$. \cite{rock1995} proposed a  model that better explains the calibration curve, it has two types of errors: an error applicable to low concentration levels and another error for larger concentrations.  Accordingly, \cite{gibbons2001weighted} consider the following model for the interlaboratory calibration data denoted by $y_{i j k}$, the $k$th measurement at the $j$th true concentration level $x_j$ in the $i$th laboratory, $i=1,2, \ldots, q, j=1,2, \ldots, r$, $k=1,2, \ldots, N_{i j}$:
\begin{equation}\label{mod_eq}
    y_{ijk} = \alpha_i + \beta_ix_j \exp(\eta_{ijk}) + \epsilon_{ijk},
\end{equation}
where $x_j$'s are the true concentration values, $j = 1,...,r$.  . We assume that
$\eta_{ijk} \widesim{i.i.d} N(0, \sigma^2_\eta), \quad \epsilon_{ijk} \widesim{i.i.d} N(0, \sigma^2_\epsilon)$  and $\eta_{ijk}$ and $\epsilon_{ijk}$ are independent. Also, the $\alpha_{i}$ and $\beta_{i}$ are the calibration parameters for the $i$th laboratory. We further assume that observations corresponding to zero concentration (i.e. $x = 0$) are available which is often the case in such experiments \cite{bhaumik2005confidence}. The $e_{i j k}$'s are the additive errors that are present primarily to explain the variability at low-level concentrations. On the other hand, the $\eta_{i j k}$'s represent proportional error at higher true concentrations, and we assume that the distributions of the proportional errors remain the same for all laboratories.   A model similar to (\ref{mod_eq}) for a single laboratory was proposed by Rocke and Lorenzato (1995) and illustrated using a gas-chromatography/mass-spectrometry example.

Note that under the model (\ref{mod_eq}) corresponding to only one laboratory (i.e., $\beta_{i}=\beta$ and $\alpha_{i}=\alpha$ ),  the data near 0 (i.e., $x \simeq 0$ ) can be used to estimate $\sigma_{e}^{2}$, and data for larger concentrations can be used  to  estimate $\sigma_{\eta}^{2}$. Model (\ref{mod_eq}) also indicates that the variability for observations at larger concentrations is mainly explained by $ e^{\eta_{i j k}} $, which follows a lognormal distribution, and observations  at low concentrations (i.e., $x \simeq 0$ ) are normally distributed, which is consistent with our experience \cite{bhaumik2005confidence}. 

\noindent\textbf{Parameter Estimation}: Multiple estimation methods have been  proposed in the literature to estimate the parameters of the model (\ref{mod_eq}). For a single interlaboratory model, \cite{rock1995} used  maximum likelihood  and applied the technique to toluene by gas-chromatography/mass-spectrometry, and cadmium by atomic absorption spectroscopy. For multiple laboratories, \cite{gibbons2001weighted} estimated model parameters by iteratively reweighted maximum marginal likelihood estimation. Performance of the method was examined via a simulation study and then it was applied to a typical interlaboratory calibration example for copper in distilled water. Employing the concept of data separation technique, i.e. separately considering the role of observations with near zero concentration and that of all other observations corresponding to higher concentrations, \cite{bhaumik2005confidence} used the  moment estimation method to estimate all model parameters. The main advantage of this method is that it gives an explicit expression for each parameter estimate, including the variance components. Inferential procedures for the calibration parameters, and confidence intervals for the  true unknown concentrations were developed using large sample properties, and implemented via  Wald  statistics. Observing some limitations of the standard likelihood based methodology in small sample size situations \cite{zhao2021accurate},  proposed a modified version of the signed log-likelihood ratio test statistic. The modified signed log-likelihood ratio test statistic provides a satisfactory approach  for developing confidence intervals for the model parameters and the true unknown concentrations. The goal of this paper is to implement the fiducial approach to achieve accurate results for  (i) determining if an analyte is present in a new sample, and  (ii)  deriving confidence bounds for a true concentration given new measurement(s).

\section{Fiducial Approach}\label{sec:fid_est}

The notion of fiducial inference is based on a fiducial quantity, introduced by  Fisher in the 1930s. Unlike Bayesian inference, it provides probabilistic statements about the unknown parameters without relying on prior distributions. The fiducial pivotal quantity is a function of random variables and corresponding observed values such that  (i) the distribution of the fiducial pivotal quantity given the data does not depend on any unknown parameters, and (ii) the observed value of the fiducial pivotal quantity obtained by replacing each random variable by the corresponding observed data is free of the nuisance parameters, and is often equal to the parameter of interest. Applications based on fiducial quantities are available in numerous articles. In particular, in the context of lognormal distributions, such applications are given in  \cite{krishnamoorthy2003lognormal},  \cite{bebu2008comparing}, and \cite{bhaumik2021}.  

\noindent {\bf Remark:} Interest in the fiducial approach was revived by the seminal works of    \cite{tsui1989} and \cite{weerahandi1993}; a fiducial quantity was referred to as a generalized pivotal quantity in the latter work.   Several later authors have developed and successfully applied the fiducial approach, referring to the fiducial quantity as a generalized pivotal quantity; see, for example, the articles by \cite{krishnamoorthy2003lognormal}, \cite{bebu2008comparing},
\cite{xu2008generalized}, \cite{bhaumik2013lognormal} and the book by \cite{kris2009}.
The fiducial pivotal quantities we have considered in our work are in the spirit of the generalized pivotal quantities considered by these authors.  We have referred to them as fiducial quantities. However,  what we have developed are approximate fiducial quantities since their implementation relies on the consistency of the maximum likelihood estimate of the model parameter $\sigma_\eta$.  In what follows, we derive fiducial quantities for each model parameter and for an unknown true concentration. 

{
\subsection{Choice of Structural Equations to Find Fiducial Pivotal Quantities and Non-Uniqueness.}
We acknowledge that, in general, fiducial inference is not invariant to the particular choice of structural equations used to construct fiducial pivotal quantities, and thus different valid constructions can lead to different fiducial distributions. A more canonical choice of structural equations is available only in special settings, most notably linear mixed-effects models under linearity and balanced-data assumptions, where the construction is guided by low-dimensional sufficient statistics and yields standard generalized confidence interval (GCI) forms (see \cite{kris2009}). Once data becomes unbalanced, even within the linear mixed-model family, such sufficient-statistic reductions are typically unavailable, and it becomes difficult to define a unique ``best" structural equation or to establish optimality properties for a particular choice. Our problem inherits this challenge because we work within a nonlinear framework and encounter unbalanced data structure across laboratories and concentrations, so no universally preferred fiducial construction becomes available. Nevertheless, our specification is not arbitrary: it is chosen to align each parameter with the empirical feature it governs and to preserve computational scalability. In what follows, we will see that for fixed $(\sigma_\eta^2,\sigma_\epsilon^2)$,  lab-specific parameters $(\alpha_i,\beta_i)$ have linear relations with the measurements $y_{ijk}$s and  transformed measurement $y_{ijk}/x_j$s ($x_j \neq 0$)  within lab $i$, which motivates us to provide structural equations based on within-lab averages leading to closed-form solutions for $\alpha_i$ and $\beta_i$.  In contrast, the variance components $(\sigma_\eta^2,\sigma_\epsilon^2)$ primarily control (i) within-lab variability across concentrations and (ii) within-concentration variability across labs; hence, we provide solutions using aggregated within-lab and within-concentration sums of squares. This choice yields an interpretable and valid fiducial construction while requiring numerical solution only for the two variance parameters, making the procedure fast and scalable as the number of labs and concentrations increases.
}

\subsection{Fiducial Quantities for  Model Parameters}
\label{subsec:pivot_mod_param}
To obtain fiducial quantities, we  rewrite the model as follows:
\begin{equation}{\label{struc_eq_mod}}
    y_{ijk} = \alpha_i + \beta_ix_j \exp(\sigma_\eta z_{ijk}^\eta) + \sigma_\epsilon z_{ijk}^\epsilon,
\end{equation}
where  $z_{ijk}^\eta \widesim{i.i.d} N(0, 1)$, $z_{ijk}^\epsilon \widesim{i.i.d} N(0, 1)$ and $z_{ijk}^\eta$, $z_{ijk}^\epsilon$ are independent. Note that (\ref{struc_eq_mod}) has observed data on the left-hand side and a function of model parameters along with random quantities on the right-hand side.  

{

\paragraph{Computational motivation and initialization.}
In principle, once a set of structural equations is specified, fiducial quantities can be obtained by jointly solving the full system of structural equations. In our setting, the structural equations in the following Sections 4.2.2-4.2.4 form a coupled nonlinear system with $(2q+2)$ unknowns (all $\{\alpha_i\}$, $\{\beta_i\}$, $\sigma_\eta^2$, and $\sigma_\epsilon^2$). Solving such a high-dimensional nonlinear system numerically at each fiducial draw can be computationally expensive and can scale poorly with the number of laboratories and concentrations. To address this, we adopt a computationally efficient construction that remains consistent with the fiducial definition by forming structural equations in blocks $\{\alpha_i\}$, $\{\beta_i\}$, and $\{\sigma_\eta^2,\sigma_\epsilon^2\}$ and, for each block, plugging in currently available fiducial quantities for the remaining parameters. This block-wise strategy naturally requires an initialization, because at the outset, fiducial quantities for the other blocks are not yet available when constructing fiducial pivotal quantities for $\{\alpha_i\}$ and $\sigma_\epsilon^2$. We therefore use the measurements at zero concentration to obtain initial fiducial pivotal quantities for $\alpha_i$ and $\sigma_\epsilon^2$, following the same motivation as \cite{bhaumik2005confidence}. We use low concentration observations (i.e. $x=0$) to form method-of-moments type estimates of these quantities. These initial fiducial pivotal quantities are then used as inputs to the subsequent structural equations for $\{\beta_i\}$ and $(\sigma_\eta^2,\sigma_\epsilon^2)$, and finally $\{\alpha_i\}$ and $\sigma_\epsilon$ are updated using the full-data structural equations (equation \eqref{pivot_alpha_i_final} and equations \eqref{struc_eq_sigma_e_eta1}-\eqref{struc_eq_sigma_e_eta2} respectively). We emphasize that this two-stage step is not an approximation; rather, it is a valid fiducial construction that enables scalable computation while allowing $\alpha_i$ and $\sigma_\epsilon$ to borrow information from the entire dataset (including the nonzero concentrations), which can improve the stability of their inference and the downstream inference for unknown concentrations.
}

\subsubsection{Initial Fiducial Quantities for $\alpha_i$ and $\sigma_\epsilon$}
\label{subsec:init_pivot}
First, we  obtain the fiducial quantities of $\alpha_i$, $i=1,...,q$, and $\sigma_\epsilon$ based on the data corresponding to $x_j = 0$. In the next subsections, we  update these fiducial quantities based on the complete data (corresponding to all concentrations $x_j$) to obtain updated fiducial quantities of $\alpha_i$, $i=1,...,q$, and $\sigma_\epsilon$. We assume $x_1=0$ and for $j=1$ the model (\ref{struc_eq_mod}) can be written as
    $y_{i1k} = \alpha_i + \sigma_\epsilon z_{i1k}^\epsilon.$
By taking an average over all replications from laboratory $i$ 
we obtain,
    $\frac{1}{N_{i1}} \sum \limits_{k=1}^{N_{i1}} y_{i1k} = \alpha_i + \sigma_\epsilon \frac{1}{N_{i1}} \sum \limits_{k=1}^{N_{i1}} z_{i1k}^\epsilon.$
Thus, we obtain,
   $\sum \limits_{k=1}^{N_{i1}} \left(y_{i1k} - \frac{1}{N_{i1}} \sum \limits_{k=1}^{N_{i1}} y_{i1k}\right)^2
    = \sigma^2_\epsilon \sum \limits_{k=1}^{N_{i1}} \left(z_{i1k}^\epsilon - \frac{1}{N_{i1}} \sum \limits_{k=1}^{N_{i1}} z_{i1k}^\epsilon \right)^2.$
By summing both sides 
over all laboratories, so as to utilize information across all labs, we get
   $\sum \limits_{i=1}^q \sum \limits_{k=1}^{N_{i1}} \left(y_{i1k} - \frac{1}{N_{i1}} \sum \limits_{k=1}^{N_{i1}} y_{i1k}\right)^2 = \sigma^2_\epsilon \sum \limits_{i=1}^q \sum \limits_{k=1}^{N_{i1}} \left(z_{i1k}^\epsilon - \frac{1}{N_{i1}} \sum \limits_{k=1}^{N_{i1}} z_{i1k}^\epsilon \right)^2$.
Thus, an initial  fiducial pivotal quantity for $\sigma_\epsilon$, say $\Tilde{\sigma}^{init}_\epsilon$,  can be obtained by solving the above equation for $\sigma_\epsilon$: 
$    \Tilde{\sigma}^{init}_\epsilon = \sqrt{\frac{\sum \limits_{i=1}^q \sum \limits_{k=1}^{N_{i1}} \left(y_{i1k} - \frac{1}{N_{i1}} \sum \limits_{k=1}^{N_{i1}} y_{i1k}\right)^2}{\sum \limits_{i=1}^q \sum \limits_{k=1}^{N_{i1}} \left(z_{i1k}^\epsilon - \frac{1}{N_{i1}} \sum \limits_{k=1}^{N_{i1}} z_{i1k}^\epsilon\right)^2}}.$
In the expression for $\Tilde{\sigma}^{init}_\epsilon$ given above, the $y_{i1k}$s are to be treated as fixed, and the random quantities are the random variable $z_{i1k}^\epsilon$. 
An initial fiducial pivotal quantity for $\alpha_i$ can be obtained as
    $\Tilde{\alpha}^{init}_i = \frac{1}{N_{i1}} \sum \limits_{k=1}^{N_{i1}} y_{i1k} - \Tilde{\sigma}^{init}_\epsilon \times \frac{1}{N_{i1}} \sum \limits_{k=1}^{N_{i1}} z_{i1k}^\epsilon.$

\medskip
\noindent {\bf Remark}. In the expression for $\Tilde{\sigma}^{init}_\epsilon$ given above, the quantity $\sum \limits_{i=1}^q \sum \limits_{k=1}^{N_{i1}} \left(z_{i1k}^\epsilon - \frac{1}{N_{i1}} \sum \limits_{k=1}^{N_{i1}} z_{i1k}^\epsilon\right)^2$ clearly has a chisquare distribution. However, while implementing the fiducial methodology, we generate the independent standard normal quantities  $z_{i1k}^\epsilon$. The reason is that these quantities also appear in later expressions; see for example equation (\ref{pivot_beta_i}). 

\subsubsection{Fiducial Quantities for $\beta_i$}

We construct a total of $q$ structural equations, one for each $\beta_i$, $i=1,...,q$. These equations are linear in each $\beta_i$, $i=1,...,q$; thus, solutions are obtained in closed forms. We use (\ref{struc_eq_mod}) to construct structural equations for each $\beta_i$, $i=1,...,q$, by averaging (\ref{struc_eq_mod}) over all replications corresponding to different concentrations,
  $\frac{1}{r} \sum \limits_{j=1}^r \frac{1}{N_{ij}} \sum \limits_{k=1}^{N_{ij}} y_{ijk} = \alpha_i + \beta_i\frac{1}{r} \sum \limits_{j=1}^r \frac{1}{N_{ij}} \sum \limits_{k=1}^{N_{ij}} x_j \exp(\sigma_\eta z_{ijk}^\eta) + \sigma_\epsilon \frac{1}{r} \sum \limits_{j=1}^r \frac{1}{N_{ij}} \sum \limits_{k=1}^{N_{ij}} z_{ijk}^\epsilon, \ i=1,...,q.$
Parameters   $\alpha_i$, $i=1,...,q$, and $\sigma^2_\epsilon$  in these structural equations are replaced  by their initial fiducial quantities, $\Tilde{\alpha}^{init}_i$, $i=1,...,q$, and  $\Tilde{\sigma}^{init}_\epsilon$ respectively. We replace $\sigma^2_\eta$ with its consistent point estimate $\hat{\sigma}^2_\eta$ which can be obtained by the  maximum likelihood  or the method of moments proposed by \cite{bhaumik2005confidence}.  Using  $\hat{\sigma}^2_\eta$, the fiducial quantities that we exhibit for the $\beta_i$, $i=1,...,q$, are only approximate fiducial quantities; however, we shall henceforth refer to the approximate fiducial quantities simply as fiducial quantities. Thus, the fiducial pivotal quantities for $\beta_i$, $i=1,...,q$, can be obtained as

\begin{equation}\label{pivot_beta_i}
    \Tilde{\beta}_i = \frac{\sum \limits_{j=1}^r \sum \limits_{k=1}^{N_{ij}} y_{ijk} - \sum \limits_{j=1}^rN_{ij}\Tilde{\alpha}^{init}_i - \Tilde{\sigma}^{init}_\epsilon \sum \limits_{j=1}^r \sum \limits_{k=1}^{N_{ij}} z_{ijk}^\epsilon}{\sum \limits_{j=1}^r \sum \limits_{k=1}^{N_{ij}} x_j \exp(\hat{\sigma}_\eta z_{ijk}^\eta)}, \quad i=1,...,q.
\end{equation}

\subsubsection{Updated Fiducial Quantities for $\alpha_i$}
We can re-write the model (\ref{struc_eq_mod}) as follows for $\{y_{ijk}: x_j \neq 0, i=1,...,q; j=1,...,r; k=1,...,N_{ij}\}$; since we assume $x_1=0$, this set is the same as  $\{y_{ijk}: i=1,...,q; j = 2, ..., r;  k=1,...,N_{ij}\}$).
\begin{equation}{\label{struc_eq_mod2}}
    \frac{y_{ijk}}{x_j} = \frac{\alpha_i}{x_j} + \beta_i \exp(\sigma_\eta z_{ijk}^\eta) + \sigma_\epsilon \frac{z_{ijk}^\epsilon}{x_j},
\end{equation}
The method for constructing equations that lead to fiducial quantities for $\alpha_i$ is similar to that for $\beta_i$ (for $i = 1,...,q$), as discussed in the previous subsection. In the following, we construct a total of $q$  equations, one for each $\alpha_i$, by summing equation (\ref{struc_eq_mod2}) over all replications at different non-zero concentrations for each lab,
  $\sum \limits_{j=2}^r \sum \limits_{k=1}^{N_{ij}} \frac{y_{ijk}}{x_j} = \alpha_i \sum \limits_{j=2}^r N_{ij}\frac{1}{x_j}  + \beta_i \sum \limits_{j=2}^r \sum \limits_{k=1}^{N_{ij}} \exp(\sigma_\eta z_{ijk}^\eta) + \sigma_\epsilon \sum \limits_{j=2}^r \sum \limits_{k=1}^{N_{ij}} \frac{z_{ijk}^\epsilon}{x_j}, \ i=1,...,q.$
Again, these are linear in each $\alpha_i$, $i=1,...,q$. We now replace $\beta_i$, $i=1,...,q$, and $\sigma^2_\epsilon$ by $\Tilde{\beta}_i$, $i=1,...,q$, and  $\Tilde{\sigma}^{init}_\epsilon$ respectively. We replace $\sigma^2_\eta$ with its consistent point estimate $\hat{\sigma}^2_\eta$. Thus, the updated fiducial pivotal quantities for $\alpha_i$, $i=1,...,q$, can be obtained as
\begin{equation}\label{pivot_alpha_i_final}
    \Tilde{\alpha}_i = \frac{\sum \limits_{j=2}^r \sum \limits_{k=1}^{N_{ij}} \frac{y_{ijk}}{x_j} - \Tilde{\beta}_i \sum \limits_{j=2}^r \sum \limits_{k=1}^{N_{ij}} \exp(\hat{\sigma}_\eta z_{ijk}^\eta) - \Tilde{\sigma}^{init}_\epsilon \sum \limits_{j=2}^r \sum \limits_{k=1}^{N_{ij}} \frac{z_{ijk}^\epsilon}{x_j}}{\sum \limits_{j=2}^r N_{ij}\frac{1}{x_j}}, \quad i=1,...,q.
\end{equation}
\subsubsection{Fiducial Pivotal Quantity for $\sigma^2_\eta$ and Updated Fiducial Pivotal Quantity $\sigma^2_\epsilon$}
In this section, we construct two structural equations and solve
them jointly for $\sigma^2_\eta$ and $\sigma^2_\epsilon$ to find their
respective fiducial quantities. As these two parameters represent variances
of random effects, appropriate sums of squares based on the observed data $\{y_{ijk}: i=1,...,q;
j=1,...,r; k=1,...,N_{ij}\}$, can be used to construct structural equations. 
From the model (\ref{mod_eq}), the variance of $y_{ijk}$ can be expressed 
as $\beta_i^2x_j^2(e^{\sigma^2_\eta} - 1)e^{\sigma^2_\eta} + \sigma^2_\epsilon$. 
Observe that for high values of $x_j$, the variability in the data inflates 
significantly. This inflates the variability in the solutions of the structural 
equations, leading to instability in root-finding algorithms and resulting in 
wide confidence intervals. To avoid this issue, we shall use the model (\ref{struc_eq_mod2}) as the 
structural equation for non-zero concentrations and (\ref{struc_eq_mod}) 
for the zero concentration. We treat $\{\frac{y_{ijk}}{x_j}: i=1,...,q; j=2,...,r; k=1,...,N_{ij}\}$ as observed 
quantities. Two structural equations are constructed based on these observed quantities--
(i) using combined within-lab variabilities for all labs (structural equation (\ref{struc_eq_sigma_e_eta1})) and
(ii) using combined within-concentration variabilities for all concentrations (structural equation (\ref{struc_eq_sigma_e_eta2})).
 Due to this reason, while constructing 
 structural equation (\ref{struc_eq_sigma_e_eta2}), we also utilized within
concentration variabilty of the obvervations corresponding to zero concentrations, which are denoted by
 $\{y_{i1k}: i=1,...,q; k=1,...,N_{i1}\}$. The structural equations for
  $\sigma^2_\eta$ and $\sigma^2_\epsilon$ can be expressed as follows:
{
\small
\begin{multline}\label{struc_eq_sigma_e_eta1}
   \sum \limits_{i=1}^q \sum \limits_{j=2}^r \sum \limits_{k=1}^{N_{ij}} \left(\frac{y_{ijk}}{x_j} - \frac{1}{r-1} \sum \limits_{j=2}^r \frac{1}{N_{ij}} \sum \limits_{k=1}^{N_{ij}} \frac{y_{ijk}}{x_j} \right)^2 =
   \sum \limits_{i=1}^q \sum \limits_{j=2}^r \sum \limits_{k=1}^{N_{ij}} \\\left(\frac{\Tilde{\alpha}_i}{x_j} + \Tilde{\beta}_i exp(\sigma_\eta z_{ijk}^\eta) + \sigma_\epsilon \frac{z_{ijk}^\epsilon}{x_j}
   - \frac{1}{r-1} \sum \limits_{j=2}^r \frac{1}{N_{ij}} \sum \limits_{k=1}^{N_{ij}} \left(\frac{\Tilde{\alpha}_i}{x_j} + \Tilde{\beta}_i exp(\sigma_\eta z_{ijk}^\eta) + \sigma_\epsilon \frac{z_{ijk}^\epsilon}{x_j}\right) \right)^2,
\end{multline}
\begin{multline}\label{struc_eq_sigma_e_eta2}
    \sum \limits_{i=1}^q \sum \limits_{k=1}^{N_{i1}} \left(y_{i1k} - \frac{1}{q}\sum \limits_{i=1}^q\frac{1}{N_{i1}} \sum \limits_{k=1}^{N_{i1}} y_{i1k} \right)^2 +
    \sum \limits_{j=2}^r \sum \limits_{i=1}^q \sum \limits_{k=1}^{N_{ij}} \left(\frac{y_{ijk}}{x_j} - \frac{1}{q} \sum \limits_{i=1}^q \frac{1}{N_{ij}}\sum \limits_{k=1}^{N_{ij}} \frac{y_{ijk}}{x_j} \right)^2 =\\
    \sum \limits_{i=1}^q \sum \limits_{k=1}^{N_{i1}} \left(\Tilde{\alpha}_i + \sigma_\epsilon z_{i1k}^\epsilon
    - \frac{1}{q}\sum \limits_{i=1}^q\frac{1}{N_{i1}} \sum \limits_{k=1}^{N_{i1}} \left(\Tilde{\alpha}_i + \sigma_\epsilon z_{i1k}^\epsilon \right) \right)^2 +
    \sum \limits_{j=2}^r \sum \limits_{i=1}^q \sum \limits_{k=1}^{N_{ij}} \\\left(\frac{\Tilde{\alpha}_i}{x_j} + \Tilde{\beta}_i exp(\sigma_\eta z_{ijk}^\eta) + \sigma_\epsilon \frac{z_{ijk}^\epsilon}{x_j}
    - \frac{1}{q} \sum \limits_{i=1}^q \frac{1}{N_{ij}} \sum \limits_{k=1}^{N_{ij}} \left(\frac{\Tilde{\alpha}_i}{x_j} + \Tilde{\beta}_i exp(\sigma_\eta z_{ijk}^\eta) + \sigma_\epsilon \frac{z_{ijk}^\epsilon}{x_j}\right) \right)^2.
\end{multline}
}

In the above structural equations, $\alpha_i$, $\beta_i$, $i=1,...,q$, are replaced by their corresponding fiducial quantities $\Tilde{\alpha}_i$, $\Tilde{\beta}_i$, $i=1,...,q$, respectively.
These structural equations are nonlinear in $\sigma_\eta^2$ and $\sigma_\epsilon^2$, and can be solved numerically 
to obtain a fiducial pivotal quantity for $\sigma_\eta^2$, and an updated fiducial pivotal quantity for $\sigma_\epsilon^2$;  denoted by $\Tilde{\sigma}_\eta^2$ and $\Tilde{\sigma}_\epsilon^2$, respectively.

{
\paragraph{Numerical solutions: Existence and uniqueness.}
For each fiducial draw, equations \eqref{struc_eq_sigma_e_eta1}-\eqref{struc_eq_sigma_e_eta2} define a two-dimensional nonlinear system in $(\sigma_\eta,\sigma_\epsilon)$ whose properties depend on the realized auxiliary variables
$\bm{z}^\eta=\{z^\eta_{ijk}\}$ and $\bm{z}^\epsilon=\{z^\epsilon_{ijk}\}$ (with the observed data $\bm{y}=\{y_{ijk}\}$ treated as fixed).
Writing the system as $E_1(\sigma_\eta,\sigma_\epsilon,\bm{z}^\eta,\bm{z}^\epsilon;\bm{y})=0$ in \eqref{struc_eq_sigma_e_eta1} and
$E_2(\sigma_\eta,\sigma_\epsilon,\bm{z}^\eta,\bm{z}^\epsilon;\bm{y})=0$ in \eqref{struc_eq_sigma_e_eta2}, we note that global existence and global uniqueness are generally difficult to guarantee for coupled nonlinear equations. However, a standard sufficient condition for local uniqueness can be stated: if a solution exists and the Jacobian matrix
$J(\sigma_\eta,\sigma_\epsilon)=\partial(E_1,E_2)/\partial(\sigma_\eta,\sigma_\epsilon)$
is nonsingular at that solution (i.e., $\det\{J(\sigma_\eta,\sigma_\epsilon)\}\neq 0$), then the solution is locally unique (by the implicit function theorem). Whether this condition holds depends on $(\bm{z}^\eta,\bm{z}^\epsilon)$ and the observed data.

In practice, since \eqref{struc_eq_sigma_e_eta1}-\eqref{struc_eq_sigma_e_eta2} involves only two parameters, we solve the system numerically using a standard root-finding routine with multiple starting values (a small $3\times 3$ grid). If a draw does not yield a solution from any starting value, we do not include that draw in the fiducial sample used for inference. If multiple solutions are returned, we select the solution that minimizes the residual sum of squares
$\sum \limits_{i,j,k}\{y_{ijk}-(\tilde{\alpha}_i+\tilde{\beta}_i x_j\exp(\sigma_\eta z^\eta_{ijk})+\sigma_\epsilon z^\epsilon_{ijk})\}^2$.
Empirically, we have observed that draws with no solution or multiple solutions become less frequent as the total sample size $\sum \limits_{i,j}N_{ij}$ increases (e.g., under the larger-sample simulation configurations such as Scenarios~1.B and~2.B).
}

\subsection{Fiducial Pivotal Quantity for an Unknown Concentration}
\label{subsec:pivot_unknown_conc}
Let $y_{ijk}^*$, $i=1,..., q^*$, $j = 1,...,r^*$, $k=1,...,N_{ij}^*$, denote observations of new samples from laboratory $i$ for an unknown concentration $\mathcal{X}_j$ at replication $k$. We assume that the laboratories where these samples are obtained are among the $q$ laboratories used to estimate the model parameters in  Model (\ref{mod_eq}). In the previous section, we obtained the fiducial quantities of all the model parameters. If we replace all the model parameters with their corresponding fiducial quantities in the structural equation (\ref{struc_eq_mod}), it  becomes  a function of the unknown concentrations. We rewrite (\ref{struc_eq_mod}) for the new samples as
\begin{equation}\label{pre_struc_eq_x_j}
    y_{ijk}^* = \alpha_i + \beta_i\mathcal{X}_j \exp(\sigma_\eta {z_{ijk}^\eta}^*) + \sigma_\epsilon {z_{ijk}^\epsilon}^*
\end{equation}
where, ${z_{ijk}^\eta}^* \widesim{i.i.d} N(0, 1)$, ${z_{ijk}^\epsilon}^* \widesim{i.i.d} N(0, 1)$ and ${z_{ijk}^\eta}^*$, ${z_{ijk}^\epsilon}^*$ are independent. A structural equation for an unknown concentration $\mathcal{X}_j$, $j=1,...,r^*$, can be obtained as follows by summing (\ref{pre_struc_eq_x_j}) over all replications and all laboratories,
    $\sum \limits_{i=1}^{q^*} \sum \limits_{k=1}^{N_{ij}^*} y_{ijk}^* = \sum \limits_{i=1}^{q^*} \sum \limits_{k=1}^{N_{ij}^*}\alpha_i + \mathcal{X}_j \sum \limits_{i=1}^{q^*} \sum \limits_{k=1}^{N_{ij}^*} \beta_i exp(\sigma_\eta {z_{ijk}^\eta}^*) + \sum \limits_{i=1}^{q^*} \sum \limits_{k=1}^{N_{ij}^*} \sigma_\epsilon {z_{ijk}^\epsilon}^*$.
We solve this for $\mathcal{X}_j$ while replacing $\alpha_i$, $\beta_i$, $i=1,...,q^*$, $\sigma_\eta$ and $\sigma_\epsilon$ by their corresponding fiducial  quantities $\Tilde{\alpha}_i$, $\Tilde{\beta}_i$, $i=1,...,q$, $\Tilde{\sigma}_\eta$ and $\Tilde{\sigma}_\epsilon$ to obtain fiducial pivotal quantity of a new concentration $\mathcal{X}_j$, $j=1,...,r^*$, as follows:
\begin{equation}\label{pivot_x_j}
    \Tilde{\mathcal{X}}_j = \frac{\sum \limits_{i=1}^{q^*} \sum \limits_{k=1}^{N_{ij}^*} y_{ijk}^* - \sum \limits_{i=1}^{q^*} \sum \limits_{k=1}^{N_{ij}^*} \Tilde{\alpha}_i - \sum \limits_{i=1}^{q^*} \sum \limits_{k=1}^{N_{ij}^*} \Tilde{\sigma}_\epsilon {z_{ijk}^\epsilon}^*}{\sum \limits_{i=1}^{q^*} \sum \limits_{k=1}^{N_{ij}^*} \Tilde{\beta}_i exp(\Tilde{\sigma}_\eta {z_{ijk}^\eta}^*)}.
\end{equation}

\begin{algorithm}[!h]
    \caption{Generating Fiducial Pivotal Quantities for Unknown Concentrations}\label{alg:fiducial}
    \begin{algorithmic}[1]
    \Require New samples $\{y^*_{ijk}: i=1,..., q^*, j = 1,...,r^*, k=1,...,N_{ij}^*\}$ with unknown true concentrations and Old samples $\{y_{ijk}: i=1,..., q, j = 1,...,r, k=1,...,N_{ij}\}$ to fit the model
    \Ensure $N$ fiducial pivotal quantities for unknown concentrations $\mathcal{X}_j$, $j=1,...,r^*$.
    \For{each replication $1,\dots,N$}
        \State Generate $z^\eta_{ijk} \overset{i.i.d.}{\sim} N(0,1)$, $z^\epsilon_{ijk} \overset{i.i.d.}{\sim} N(0,1)$, for $i=1,...,q$, $j=1,...,r$, $k=1,...,N_{ij}$.
        \State Generate initial fiducial quantities for $\sigma_\epsilon$ and $\alpha_i$, as $\tilde{\sigma}^{init}_\epsilon$ and $\tilde{\alpha}^{init}_i$, respectively, for each $i=1,\dots,q$ using the method in section \ref{subsec:init_pivot}.
        \State Generate fiducial quantities for $\beta_i$, $i=1,\dots,q$, using expression (\ref{pivot_beta_i}), and update fiducial quantities for $\alpha_i$ using expression (\ref{pivot_alpha_i_final}). In this step, a point estimate of $\sigma_\eta$ is needed. 
        \State Generate fiducial pivotal quantity for $\sigma_\eta$ and updated fiducial pivotal quantity for $\sigma_\epsilon$ by solving equations (\ref{struc_eq_sigma_e_eta1}) and (\ref{struc_eq_sigma_e_eta2}) simultaneously.
        \State Generate ${z^\eta_{ijk}}^* \overset{i.i.d.}{\sim} N(0,1)$, ${z^\epsilon_{ijk}}^* \overset{i.i.d.}{\sim} N(0,1)$ for $i=1,...,q^*$, $k=1,...,N_{ij}^*$.
        \State Obtain fiducial pivotal quantity of unknown concentration $\mathcal{X}_j$ using expression (\ref{pivot_x_j}) for each $j=1,...,r^*$.
    \EndFor
    \State \Return $N$ fiducial pivotal quantities for each unknown concentration $\mathcal{X}_j$, $j=1,...,r^*$ from step 7.
    \end{algorithmic}
\end{algorithm}

\subsection{Fiducial Confidence Interval and Fiducial Point Estimate}

So far we have discussed a method of constructing fiducial
quantities of the parameters under the model in Subsection \ref{subsec:pivot_mod_param}.
Furthermore, in subsection \ref{subsec:pivot_unknown_conc}, we have discussed how to obtain
fiducial quantities for the unknown true concentrations corresponding to new measurements utilizing the fiducial 
quantities obtained in subsection \ref{subsec:pivot_mod_param}. In this section, we
conclude our discussion on fiducial inference for the unknown true concentrations by discussing
how to utilize these fiducial quantities to obtain point and interval estimates of the
unknown concentrations.

Since fiducial quantities are functions of the observed data and random quantities, whose distributions are free of any unknown parameters, samples from the fiducial quantities can be obtained simply by drawing samples from the corresponding distributions. These samples can then be used to estimate the density of the fiducial distribution for each parameter. Once the density is estimated, it serves as a basis for determining both point and interval estimates for each parameter. By examining the estimated density, we can derive point estimates such as the mode, and the highest density intervals, which offer a measure of uncertainty around these estimates. In this work, we focus on the point and interval estimation of the unknown concentration from new samples $\{y^*_{ijk}: i=1,..., q^*, j = 1,...,r^*, k=1,...,N_{ij}^*\}$, where the fiducial quantities of the model parameters have been obtained from the earlier sample $\{y_{ijk}: i=1,..., q, j = 1,...,r, k=1,...,N_{ij}\}$, corresponding to known concentrations. For an unknown concentration $\mathcal{X}_j$ (for any $j=1,..,r^*$), one sample from its fiducial distribution can be drawn sequentially based on the discussion in sections \ref{subsec:pivot_mod_param} and \ref{subsec:pivot_unknown_conc} using algorithm \ref{alg:fiducial}. One can generate $N$ fiducial quantities of each unknown concentration by replicating these steps $N$ times. Let us denote the density of the fiducial distribution of the unknown concentration $\mathcal{X}_j$ as $d_{\mathcal{X}_j}(.)$ which can be estimated accurately from the $N$ generated fiducial quantities. In the following, we  use the estimated 95\% fiducial highest density interval as the interval estimate and the fiducial mode as the point estimate of  $\mathcal{X}_j$.

\noindent \textbf{100$(1-\alpha)$\% Fiducial Highest Density Interval (HDI).} The 100$(1-\alpha)$\% fiducial HDI for $\mathcal{X}_j$ is the interval $[L, U]$ such that: $\int \limits_L^U d_{\mathcal{X}_j}(x) \delta x = 1-\alpha$ and minimizes the interval length $(U - L)$.  The endpoints $L$ and $U$ are determined such that $d_{\mathcal{X}_j}(L) = d_{\mathcal{X}_j}(U)$ ensuring that the interval captures the highest density region.\\
\textbf{Fiducial Mode.} The fiducial mode $\hat{\mathcal{X}_j}$ is the value of $\mathcal{X}_j$ that maximizes the fiducial density function; i.e.,  $\hat{\mathcal{X}_j} = \arg\max_{x} d_{\mathcal{X}_j}(x)$. This point estimate $\hat{\mathcal{X}_j}$ represents the value of $\mathcal{X}_j$ with the highest probability density under the fiducial distribution.

\section{Simulation Studies}
To illustrate some key features of the proposed confidence interval for a true unknown concentration, we shall use a few different simulation scenarios. For each simulation scenario, training and test samples (as described in Section 5.1) are generated using two distinct configurations. We label the two configurations for training sample generation as ``A" and ``B," and the two configurations for test sample generation as ``1" and ``2." This results in a total of four simulation scenarios: 1.A, 2.A, 1.B, and 2.B. The difference between configurations A and B lies in the generation of the training samples. Configuration A includes 3 true concentrations and 3 laboratories, while configuration B uses 9 true concentrations and 10 laboratories. On the other hand, configurations 1 and 2 differ in how the test samples are generated. In configuration 1, multiple replications are drawn from a single laboratory to estimate the unknown concentration, whereas configuration 2 involves a single replication from multiple laboratories. In all scenarios, we compare the performance of our proposed fiducial method for interval estimation with several alternative methods:  the parametric bootstrap, the adjusted likelihood ratio test (LRT) statistics proposed by \cite{zhao2021accurate}, Wald-type large-sample method developed by \cite{bhaumik2005confidence} using Method of Moment Estimator (MME), and Wald-type method based on the MLE.
\subsection{Simulation Setup} \textbf{Details of training samples for simulation Scenarios 1.A and 2.A.} Data were simulated for three concentrations (0, 10, 30) $\mu \mathrm{g} / \mathrm{L}$ with $N_{i j}=5$ replicates per concentration nested within $q=3$ laboratories.  Values of the model (\ref{mod_eq}) parameters were set at: $\alpha_i=1$, $\beta_i=1, \sigma_{e}=1$, and $\sigma_{\eta}=0.1$. This setup corresponds to a small sample size as only three laboratories and three concentrations per laboratory have been considered.\\
\textbf{Details of training samples for simulation Scenarios 1.B and 2.B.}  Data were simulated with a moderate sample size for nine concentrations (0, 5, 10, 15, 20, 25, 30, 35, 40)  $\mu \mathrm{g} / \mathrm{L}$ with $N_{i j}=5$ replicates per concentration nested within $q=10$ laboratories. Values of the model parameters were set at: $\alpha_i=0$, $\beta_i=1, \sigma_{e}=1$, and $\sigma_{\eta}=0.1$. \\ 
\textbf{Details of test samples for simulation Scenarios 1.A and 1.B.}  Samples for an unknown concentration consist of 5 replications from one laboratory (without loss of generality, say Laboratory 1); with $q^*=1$, $N_{i j}^*=5$. For Scenario 1.A,  true values of unknown concentrations are (5, 20, 50). In Scenario 1.B,  true values of unknown concentrations are (0, 5, 10, 15, 20, 25, 30, 35, 40, 50) $\mu \mathrm{g} / \mathrm{L}$.\\
\textbf{Details of test samples for simulation Scenarios 2.A and 2.B.} 
The samples for an unknown concentration consist of a single replication from multiple laboratories. In scenario 2.A, we use three laboratories in the training sample, i.e., $q^*=3$, $N_{i j}^*=1$, while, in scenario 2.B,  $q^*=6$, $N_{i j}^*=1$. The true values of unknown concentrations for 2.A and 2.B are the same as 1.A and 1.B, respectively. 
\subsection{Simulation Results}
\textbf{Interval Estimation}: The simulation results for each scenario are based on 5,000 simulated datasets. For each dataset, the proposed fiducial method estimates the unknown concentration using 10,000 samples generated from the fiducial pivotal quantity. Similarly, 10,000 bootstrap samples are utilized in both the parametric bootstrap and LRT-based methods for comparison. Notably, the LRT-based method is computationally more intensive than the parametric bootstrap method. However, the proposed fiducial method provides significantly more computational efficiency compared to both approaches. The simulation results for interval estimation are presented in Tables \ref{tab:sim_1_A}, \ref{tab:sim_1_B}, \ref{tab:sim_2_A}, \ref{tab:sim_2_B}. For simulation scenarios 1.B and 2.B (Tables \ref{tab:sim_1_B} and \ref{tab:sim_2_B}), results for concentrations 0, 15, 30, and 50 are presented in this document (Complete tables can be found in Supplementary Material).  The performance of the interval estimate is evaluated based on both the average width of the 95\% confidence interval and its coverage probability.
\begin{figure}
    \centering
    \includegraphics[height = 7cm, width = 18cm]{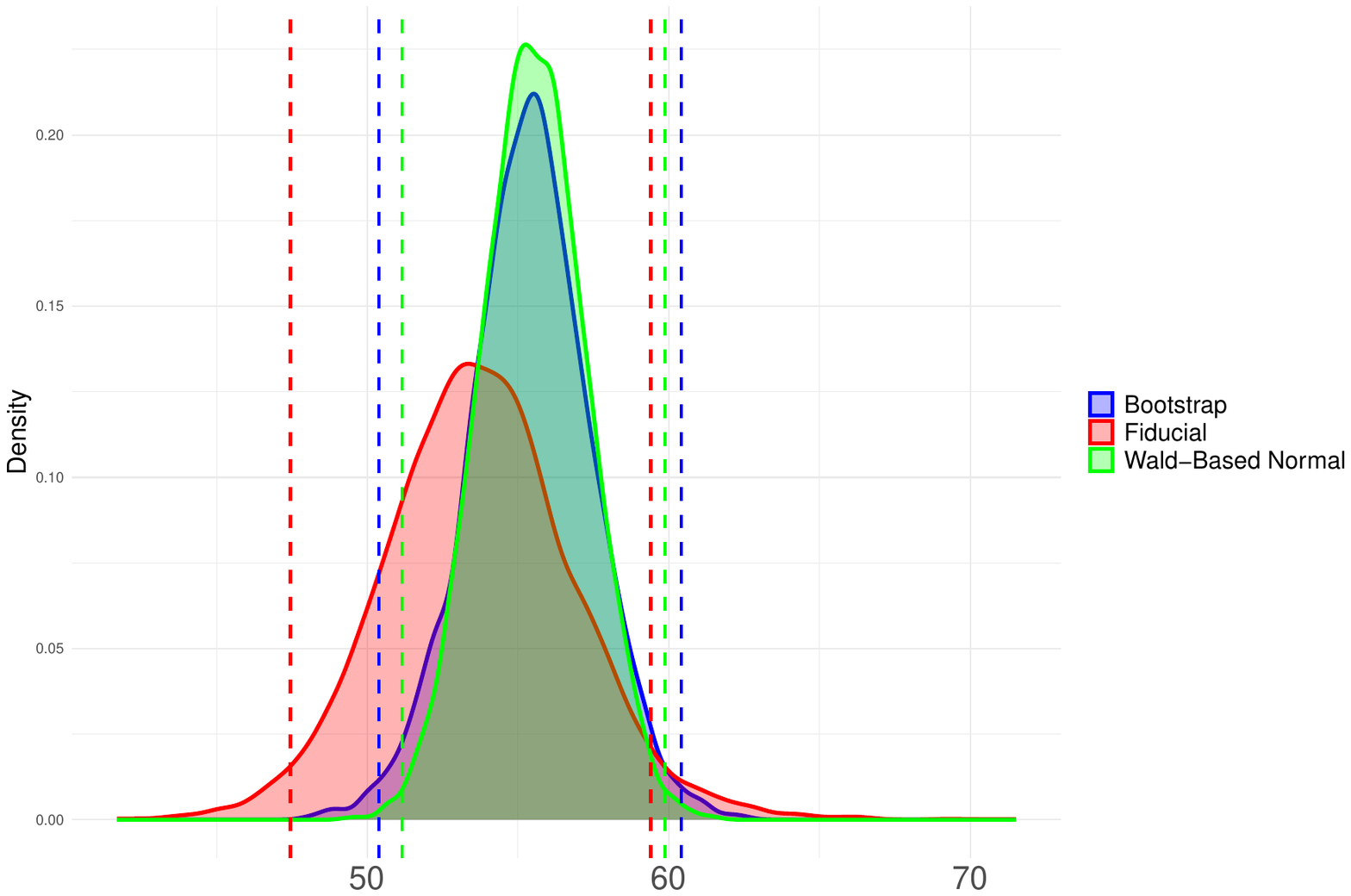}
    \caption{\footnotesize Density plots of the fiducial distribution, bootstrap distribution and Wald-Based normal distribution for an unknown concentration (true value is 50) for a sample from simulation setting 2.A where the MLE is 55.6. Vertical dashed lines shows 95\% confidence intervals. The figure highlights the fact that only the fiducial interval includes the true value: a conclusion that agrees with the poor coverage probabilities of the other intervals, as reported in this section. { Note that in the overlaid contour/density display, apparent colour changes at intersections can occur as a rendering artifact of overlapping densities and do not represent an additional method.}}
    \label{fig:density_plot}
\end{figure}
\begin{table}[h]
\caption{Estimated coverage probabilities and expected widths of the 95\% confidence intervals for the true unknown concentration: Simulation Scenario 1.A}
\centering
\vspace{.4cm}
\renewcommand{\arraystretch}{0.7}
\fontsize{9}{10}\selectfont
\begin{tabular}{|c|c|c|c|c|c|}
\hline 
True & \multicolumn{5}{|c|}{Average 95\% Confidence Limits [Average Width]}\\
\cline{2-6}
Conc. & Fiducial & Bootstrap & LRT & Wald MLE & Wald MME \\
\hline
5 & (3.8, 6.3) [2.5] & (4.0, 6.4) [2.4] & (3.7, 6.3) [2.6] & (4.0, 6.2) [2.2] & (2.8, 7.2) [4.4]\\
20 & (17.6, 22.7) [5.1] & (17.6, 22.4) [4.8] & (17.8, 22.8) [5.0] & (17.8, 22.3) [4.5] & (15.6, 24.6) [9.0]\\
50 & (44.2, 56.5) [12.3] & (43.3, 55.5) [12.2] & (44.4, 56.5) [12.1] & (44.4, 55.3) [10.9] & (40.1, 60.4) [20.3]\\
\hline
True & \multicolumn{5}{|c|}{Coverage Probability}\\
\cline{2-6}
Conc. & Fiducial & Bootstrap & LRT & Wald MLE & Wald MME\\
\hline
5 & 0.949 & 0.931 & 0.941 & 0.931 & 1.000\\
20 & 0.952 & 0.938 & 0.889 & 0.921 & 1.000\\
50 & 0.949 & 0.918 & 0.915 & 0.911 & 0.994\\
\hline
\end{tabular}
\label{tab:sim_1_A}
\end{table}
\begin{table}[h]
\caption{Simulation Scenario 1.B: Estimated coverage probabilities and expected widths of the 95\% confidence intervals for the true unknown concentrations 0, 15, 30, 50.}
\centering
\vspace{.4cm}
\renewcommand{\arraystretch}{1}
\fontsize{9}{10}\selectfont
\begin{tabular}{|c|c|c|c|c|c|c|c|c|}
\hline 
True & \multicolumn{4}{|c|}{Average 95\% Limits [Average Width]} & \multicolumn{4}{|c|}{Coverage Probability}\\
\cline{2-9}
Conc. & Fiducial & Bootstrap & Wald & Wald & Fiducial & Bootstrap & Wald & Wald\\
& & & MLE & MME & & & MLE & MME\\
\hline
0 & (0, 1.1) [1.1] & (0, 1.5) [0.9] & (0, 1.3) [1.3] & (0, 1.9) [1.9] & 0.948 & 0.814 & 0.900 & 0.996\\
15 & (13.4, 16.7) [3.3] & (13.6, 16.7) [3.1] & (13.4, 16.7) [3.3] & (11.3, 18.6) [7.3] & 0.949 & 0.931 & 0.943 & 1.000\\
30 & (27.1, 33.0) [5.9] & (26.8, 32.5) [5.7] & (26.9, 32.7) [5.8] & (23.6, 36.3) [12.7] & 0.947 & 0.914 & 0.942 & 1.000\\
50 & (45.3, 55.2) [9.9] & (44.2, 53.6) [9.4] & (44.9, 54.4) [9.5] & (39.8, 60.4) [20.6] & 0.949 & 0.899 & 0.935 & 1.000\\
\hline
\end{tabular}
\label{tab:sim_1_B}
\end{table}
As already noted, these methods include  (i) the parametric bootstrap, (ii) the modified LRT statistic proposed by \cite{zhao2021accurate}, (iii) Wald-type large-sample method developed by \cite{bhaumik2005confidence} using  MME, and (iv) Wald-type method based on the MLE. Aside from the simulation scenario 1.A, the modified LRT approach is not viewed as a competing method as it (a)  demands extremely high computational time (discussed later in this section) and (b) 
 has a relatively poor performance compared to the proposed method shown in Table \ref{tab:sim_1_A}. Similarly, the Wald-type method  based on the MME  is  not considered a competitor either (expect scenario 1.A), due to its  conservative nature shown in Table \ref{tab:sim_1_A}. In the following we summarize the results of the simulation scenarios.

When the size of the training sample is small (as seen in simulation Scenarios 1.A and 2.A), the coverage probability of the proposed 95\% fiducial confidence interval is approximately 0.95. In contrast, other methods such as the bootstrap and Wald, exhibit coverage probabilities that are either significantly smaller than  0.95 or very close to 1 (see Tables \ref{tab:sim_1_A} and \ref{tab:sim_2_A}), and the expected confidence widths correspondingly reflect the behavior of the coverage probabilities.  As the training sample size increases from small to moderate (Scenarios 1.B and 2.B), the performance of the Wald intervals based on the MLE improves, as expected. Nevertheless, even in these scenarios, the coverage probability of the fiducial interval remains the most accurate. In fact  we see that the coverage probability associated with the parametric bootstrap solution can be significantly lower than the assumed confidence level. We have provided Figure \ref{fig:density_plot} for illustration for such lower coverage of other intervals.

\noindent \textbf{Point Estimation}: The same simulated datasets are used for evaluate the results for the point estimation. The performance of the point estimate is evaluated based on relative bias, relative absolute bias, and relative root mean squared error, which are defined in the supplementary material. We compare the proposed fiducial method to the MLE. The simulation results for point estimation are displayed in Table \ref{tab:sim_1_point_est} (complete tables with all concentrations are in supplementary). The numerical results indicate that the fiducial method demonstrates less bias than the MLE across all simulation scenarios. This feature is particularly pronounced at low concentrations, such as 0 and 5, where the relative bias of the MLE is significantly larger than that of the fiducial method (see Table \ref{tab:sim_1_point_est}). Furthermore, aside from exhibiting less bias, the fiducial method shows comparable performance with the MLE in terms of both absolute relative bias and relative root mean squared error.
\begin{table}[h]
\caption{Estimated coverage probabilities and expected widths of the 95\% confidence intervals for the true unknown concentration: Simulation Scenario 2.A}
\centering
\vspace{.4cm}
\renewcommand{\arraystretch}{0.8}
\fontsize{9}{10}\selectfont
\begin{tabular}{|c|c|c|c|c|c|c|}
\hline 
True & \multicolumn{3}{|c|}{Average 95\% Limits [Average Width]} & \multicolumn{3}{|c|}{Coverage Probability}\\
\cline{2-7}
Conc. & Fiducial & Bootstrap & Wald & Fiducial & Bootstrap & Wald\\
& & & MLE & & & MLE\\
\hline
5 & (3.6, 6.4) [2.8] & (3.8, 6.4) [2.6] & (3.8, 6.2) [2.2] & 0.949 & 0.939 & 0.918\\
20 & (17.3, 22.9) [5.6] & (17.3, 22.6) [5.3] & (17.7, 22.4) [4.7] & 0.953 & 0.938 & 0.927\\
50 & (43.8, 56.7) [12.9] & (43.2, 55.8) [12.6] & (44.4, 55.6) [11.2] & 0.947 & 0.901 & 0.895\\
\hline
\end{tabular}
\label{tab:sim_2_A}
\end{table}

\begin{table}[h]
\caption{Simulation Scenario 2.B: Estimated coverage probabilities and expected widths of the 95\% confidence intervals for the true unknown concentrations 0, 15, 30, 50.}
\centering
\vspace{.4cm}
\renewcommand{\arraystretch}{0.8}
\fontsize{9}{10}\selectfont
\begin{tabular}{|c|c|c|c|c|c|c|}
\hline 
True & \multicolumn{3}{|c|}{Average 95\% Limits [Average Width]} & \multicolumn{3}{|c|}{Coverage Probability}\\
\cline{2-7}
Conc. & Fiducial & Bootstrap & Wald & Fiducial & Bootstrap & Wald\\
& & & MLE & & & MLE\\
\hline
0 & (0, 0.9) [0.9] & (0, 1.1) [1.1] & (0, 1.0) [1.0] & 0.952 & 0.849 & 0.936\\
15 & (13.6, 16.5) [2.9] & (13.7, 16.4) [2.7] & (13.6, 16.5) [2.9] & 0.948 & 0.918 & 0.938\\
30 & (27.5, 32.6) [5.1] & (27.4, 32.2) [4.8] & (27.4, 32.4) [5.0] & 0.946 & 0.932 & 0.942\\
50 & (46.0, 54.2) [8.2] & (45.4, 53.3) [7.9] & (45.7, 53.7) [8.0] & 0.948 & 0.912 & 0.932\\
\hline
\end{tabular}
\label{tab:sim_2_B}
\end{table}
\begin{table}[h]
\caption{Comparison of point estimates by the fiducial method and maximum likelihood for simulation Scenarios. (Fid: Fidicial Mode, MLE: Maximum Likelihood Estimator, Rel. Bias: Relative Bias, Rel. Abs. Bias: Relative Absolute Bias, Rel. RMSE: Relative Root Mean Square Error. For zero concentration Bias, Abs. Bias, RMSE are computed)}
\centering
\vspace{.4cm}
\renewcommand{\arraystretch}{0.8}
\fontsize{9}{10}\selectfont
\begin{tabular}{|c|c|c|c|c|c|c|c|c|c|c|c|c|c|}
\hline
\multicolumn{7}{|c|}{Scenario 1.A} & \multicolumn{7}{|c|}{Scenario 1.B}\\
\hline
True & \multicolumn{2}{|c|}{Rel.} & \multicolumn{2}{|c|}{Rel.} & \multicolumn{2}{|c|}{Rel.} & True & \multicolumn{2}{|c|}{Rel.} & \multicolumn{2}{|c|}{Rel.} & \multicolumn{2}{|c|}{Rel.}\\
Conc. & \multicolumn{2}{|c|}{Bias} & \multicolumn{2}{|c|}{Abs. Bias} & \multicolumn{2}{|c|}{RMSE} & Conc. & \multicolumn{2}{|c|}{Bias} & \multicolumn{2}{|c|}{Abs. Bias} & \multicolumn{2}{|c|}{RMSE}\\
& \multicolumn{2}{|c|}{($\times 10^2$)} & \multicolumn{2}{|c|}{($\times 10^2$)} & \multicolumn{2}{|c|}{($\times 10^2$)} & & \multicolumn{2}{|c|}{($\times 10^2$)} & \multicolumn{2}{|c|}{($\times 10^2$)} & \multicolumn{2}{|c|}{($\times 10^2$)}\\
\hline
& Fid & MLE & Fid & MLE & Fid & MLE & & Fid & MLE & Fid & MLE & Fid & MLE\\
\hline
5 & -0.22 & 2.53 & 9.8 & 9.7 & 12 & 12 & 0 & -1.54 & 24 & 46 & 51 & 59 & 63\\
20 & -0.01 & 0.20 & 5.0 & 5.0 & 6.4 & 6.3 & 5 & -0.58 & 4.56 & 9.0 & 9.5 & 11 & 12\\
50 & 0.01 & -0.31 & 4.8 & 4.7 & 5.9 & 5.9 & 10 & 0.50 & 1.46 & 5.8 & 5.7 & 7.2 & 7.1\\
\hline
\end{tabular}
\begin{tabular}{|c|c|c|c|c|c|c|c|c|c|c|c|c|c|}
\hline
\multicolumn{7}{|c|}{Scenario 2.A} & \multicolumn{7}{|c|}{Scenario 2.B}\\
\hline
True & \multicolumn{2}{|c|}{Rel.} & \multicolumn{2}{|c|}{Rel.} & \multicolumn{2}{|c|}{Rel.} & True & \multicolumn{2}{|c|}{Rel.} & \multicolumn{2}{|c|}{Rel.} & \multicolumn{2}{|c|}{Rel.}\\
Conc. & \multicolumn{2}{|c|}{Bias} & \multicolumn{2}{|c|}{Abs. Bias} & \multicolumn{2}{|c|}{RMSE} & Conc. & \multicolumn{2}{|c|}{Bias} & \multicolumn{2}{|c|}{Abs. Bias} & \multicolumn{2}{|c|}{RMSE}\\
& \multicolumn{2}{|c|}{($\times 10^2$)} & \multicolumn{2}{|c|}{($\times 10^2$)} & \multicolumn{2}{|c|}{($\times 10^2$)} & & \multicolumn{2}{|c|}{($\times 10^2$)} & \multicolumn{2}{|c|}{($\times 10^2$)} & \multicolumn{2}{|c|}{($\times 10^2$)}\\
\hline
& Fid & MLE & Fid & MLE & Fid & MLE & & Fid & MLE & Fid & MLE & Fid & MLE\\
\hline
5 & -0.84 & 1.12 & 10 & 10 & 13 & 13 & 0 & -0.76 & 15 & 34 & 37 & 42 & 45\\
20 & -0.06 & 0.19 & 5.3 & 5.3 & 6.7 & 6.7 & 5 & -0.12 & 3.27 & 7.5 & 7.8 & 9.5 & 10\\
50 & 0.00 & -0.10 & 5.5 & 5.5 & 6.9 & 6.8 & 10 & -0.09 & 1.27 & 4.5 & 4.5 & 5.7 & 5.7\\
\hline
\end{tabular}
\label{tab:sim_1_point_est}
\end{table}

\subsection{Computational Efficiency of the Fiducial Approach} Computational efficiency should be an important factor when developing new methods, especially for interdisciplinary applications, as these methods may be integrated into publicly available software and will be adopted by researchers across disciplines. Thus, in addition to demonstrating the superiority of the fiducial confidence intervals, we highlight its computational advantage over the other available methods. The fiducial approach entails drawing fiducial samples; for this, we have to numerically solve only two equations (for $\sigma_\eta$ and $\sigma_\epsilon$) since closed-form expressions are available for the fiducial pivotal quantities for all other model parameters ($\alpha_i, \beta_i, i=1,...,q$), as well as the unknown concentrations $\mathcal{X}_j, j=1,...,r$. Thus, obtaining a fiducial sample is computationally very cheap. All other approaches, including the bootstrap, require extensive maximum likelihood calculations, including numerical integration and numerical optimization of the integrated likelihood. This can be computationally extensive compared to the proposed fiducial method, especially  when the data come from multiple laboratories. The computation is especially burdensome for the modified likelihood ratio method due to \cite{zhao2021accurate} since it involves likelihood maximization and bootstrap sampling at several stages. 

Let us denote the computation time for the bootstrap method as `$t$'. Now, coming into Zhao's LRT, it needs a large number of bootstrap samples and computation of constrained and unconstrained MLEs to obtain a single value of the standardized LRT statistics. Note that this computational time is ‘$2t$’ (each `$t$' for constrained and unconstrained MLE).  Further, obtaining each limit of the confidence interval requires the generation of ‘$k$’ values of the standardized LRT. Thus, a confidence interval of a single unknown concentration requires computation time ‘$4kt$’. To achieve decent accuracy, a value of ‘$k$' is considered at least 50. Additionally, these samples can not be used again to obtain intervals for another unknown concentration, even if the samples come from the same lab. Now, if there are 10 unknown concentrations of interest, the computation time will increase simply 10 times. The proposed fiducial method can utilize the same fiducial pivotal quantity of the model parameters to obtain fiducial pivotal quantities for all unknown concentrations. In summary, to obtain intervals for 10 unknown concentrations, if the computation time for bootstrap is ‘$t$’, LRT takes ‘$2000t$’, whereas fiducial samples are take less than ‘$t$’,  as a fiducial sample is computationally much cheaper than a bootstrap sample.
The above discussion shows the computational extensiveness of Zhao's method.  It becomes computationally impractical to evaluate its performance in simulation cases with multiple laboratories and many unknown concentrations. This is one of the reasons why it is not being used in simulation cases apart from simulation scenario 1.A.

{
\paragraph{Example of wall-clock runtimes.}
To complement the computational comparison in terms of $t$, we also report wall-clock runtimes. All computations were carried out on a laptop with Intel(R) Core(TM) i7-10510U CPU @ 1.80GHz (2.30GHz) and 16GB RAM. Using 1000 resamples/samples, the average bootstrap runtime is approximately $t\approx 320$ seconds in the smaller-sample scenarios (1.A and 2.A), and $t\approx 3680$ seconds in the larger-sample scenarios (1.B and 2.B). In contrast, the proposed fiducial method requires approximately 241 seconds on average in the larger-sample scenarios using 1000 fiducial samples. Translating the scaling argument above, for 10 unknown concentrations the modified LRT method of \cite{zhao2021accurate} would require approximately $2000t$, which corresponds to about $7.36\times 10^6$ seconds (approximately 85 days) when $t=3680$ seconds in the larger-sample setting, illustrating its computational impracticality for the multi-laboratory/multiple-unknown regime studied here.
}

\section{Data Analysis}\label{sec:data_analysis}
We report the results from the analysis of both the cadmium and copper datasets mentioned earlier. One measurement from each laboratory was selected as the test sample for each concentration, while the remaining data were used to train the model. Since we are working with real-world datasets, it is essential to first evaluate the adequacy of the Rocke-Lorenzato model specified in (\ref{mod_eq}) before performing the data analysis to estimate the unknown concentrations. To this end, the training data was used to fit the Rocke-Lorenzato model (\ref{mod_eq}). Subsequently, 95\% upper and lower calibration curves were constructed based on the fitted model,  the majority of observed measurements fall within the specified limits, supporting the adequacy of the Rocke-Lorenzato model (\ref{mod_eq}) for these datasets. A detailed goodness-of-fit assessments for the Rocke–Lorenzato calibration model on these datasets is presented in the supplementary material.

After assessing the model fit, the test samples with unknown concentrations are now analyzed, leading to the construction of confidence intervals and point estimates for the unknown concentrations. Table \ref{tab:data_results} presents point estimates calculated using both the fiducial method and the MLE for each concentration. Accompanying these point estimates are 95\% confidence intervals derived from the proposed fiducial method in Section \ref{sec:fid_est}, as well as those obtained using the bootstrap, Wald methods based on MLE and LRT by \cite{zhao2021accurate}.
\begin{table}[h]
\caption{Results for Copper and Cadmium data. Each data consists five replicates for each concentration in every laboratory. The first four replicates were used to train the data and the last replicate was used to estimate the concentration. }
\centering
\vspace{.4cm}
\renewcommand{\arraystretch}{0.9}
\fontsize{10}{10}\selectfont
\begin{tabular}{|c|c|c|c|c|c|c|}
\hline
\multicolumn{7}{|c|}{Cadmium Data}\\
\hline
True & \multicolumn{2}{|c|}{Point Estimate} & \multicolumn{4}{|c|}{95\% Confidence Interval}\\
\cline{2-7}
Conc. & Fiducial & MLE & Fiducial & Bootstrap & MLE & LRT\\
\hline
0 & 1.40 & 1.51 & (0, 5.6) & (0, 5.4) & (0, 5.0) & (0, 5.8)\\
20 & 20.5 & 19.4 & (17, 24) & (16, 24) & (16, 23) & (16, 24)\\
100 & 103.2 & 102.8 & (91, 107) & (96, 110) & (96, 110) & (95, 112)\\
\hline
\multicolumn{7}{|c|}{Copper Data}\\
\hline
True & \multicolumn{2}{|c|}{Point Estimate} & \multicolumn{4}{|c|}{95\% Confidence Interval}\\
\cline{2-7}
Conc. & Fiducial & MLE & Fiducial & Bootstrap & MLE & LRT\\
\hline
0 & 0.14 & 0.20 & (0, 2.7) & (0, 2.1) & (0, 2.1) & (0, 3)\\
2 & 2.6 & 2.6 & (0.6, 5.0) & (1.1, 4.7) & (0.7, 4.5) & (0, 6)\\
10 & 9.1 & 8.9 & (6.5, 11.5) & (6.8, 11.0) & (7.0, 10.8) & (6.1, 11.8)\\
50 & 47.3 & 47.1 & (42.6, 51.5) & (44.7, 49.8) & (44.7, 49.5) & (38.9, 48.5)\\
200 & 197.1 & 195.9 & (181, 208) & (189, 205) & (189, 203) & (186, 204)\\
\hline
\end{tabular}
\label{tab:data_results}
\end{table}
Notably, we observed in the simulation studies that unlike fiducial confidence intervals, which exhibit highly accurate coverage probabilities, the bootstrap, Wald MLE, and Zhao's LRT methods often yield coverage probabilities that fall significantly below the intended confidence level. Clearly, this has implications for the application of these methodologies.  For example, for the copper data, when the true concentration is 50, both the bootstrap and Wald MLE methods yield upper limits of the 95\% interval that fall below 50. This demonstrates the potential unreliability of these methods.

\section{Discussion}

The fiducial approach developed in this article is unique in that it allows estimates for a single laboratory or multiple laboratories analyzing split samples from the same environmental source to borrow strength from other laboratories for which calibration data using the same analytic methodology are available. The distinct advantage of this approach is that both the point estimate and confidence interval are more precise, and computationally it is easy to implement. The approach presented in this paper addresses a crucial challenge in analytical chemistry, namely the uncertainty in chemical measurements. Uncertainty is frequently underestimated in this field, leading to the erroneous belief that measurement estimates reflect true values. The techniques devised in this study facilitate the estimation of uncertainty related to the actual true concentration upon which these measurements are predicated. Given that measurements are typically collected from various instruments or analyses, and often across multiple laboratories, it is essential to consider these sources of variability when evaluating the overall uncertainty in the analytical process. Therefore, these variances should be factored into calculations that define the bounds for the underlying true concentration.
In practical applications, when assessing a measurement against a regulatory threshold limit, or compliance standard (referred to as RT), we can compute a confidence interval for the true unknown concentration \(X\). An exceedance is declared only if the lower confidence limit (LCL) exceeds the RT, providing 100\((1 - \alpha)\%\) confidence that the RT has been surpassed.  Thus, determination of the  LCL with the prespecified confidence level is extremely important for comparisons to the RT, and our fiducial approach provides an LCL for this purpose. 

{
 Throughout this paper, we use the term ``borrowing strength" in a purely descriptive sense to denote information sharing induced by the joint fiducial construction, and not to indicate an additional hierarchical (random-effects) model for the laboratory-specific calibration parameters. In the proposed fiducial method, borrowing strength occurs because the variance components $(\sigma_\eta^2,\sigma_\epsilon^2)$ are common across laboratories and are estimated by pooling information over all laboratories and concentrations via equations \eqref{struc_eq_sigma_e_eta1}-\eqref{struc_eq_sigma_e_eta2}. This pooled uncertainty, together with  lab-specific parameters $(\alpha_i,\beta_i)$, is then propagated into inference on unknown concentrations. As a result, concentration intervals are stabilized relative to analyses that treat each laboratory separately, while still allowing the calibration relationships to differ across laboratories through $(\alpha_i,\beta_i)$.
}

Two additional noteworthy applications of this method are worth mentioning. First, it can be employed to estimate a detection limit, defined as the concentration threshold where true values exceed zero (see \cite{currie1968limits, danzer1998guidelines, gibbons1996groundwater}). In this context, we identify the minimum value of \(X\) such that the LCL is greater than zero. Secondly, this method can aid in determining a quantification limit (see \cite{currie1968limits, gibbons1992practical}). The goal in this case is to identify the true concentration for which the confidence interval is sufficiently narrow to ensure a reliable quantitative measurement. \cite{currie1968limits} defined the ``determination limit" as the true concentration at which the percent relative standard deviation (RSD) is 10\%. We propose redefining the quantification limit as the value of \(X\) where the interval length is 10\% of the point estimate of \(X\) given \(Y\). Previous methodologies primarily concentrated on single-laboratory confidence bounds on \(Y\) given \(X\) (see \cite{gibbons1996groundwater, gibbons2001weighted}), rather than directly computing bounds on \(X\) based on single or multiple laboratories, as this paper demonstrates. It is crucial to recognize that achieving a 10\% RSD may not always be attainable, contingent upon the variance (\(\sigma_{\eta}\)).

\section*{Acknowledgements}
The authors are grateful to two reviewers for their helpful comments and positive feedback. Their comments resulted in the clarification of some of the ideas and improved presentation of the results.

\bigskip
\begin{center}
{\large\bf SUPPLEMENTARY MATERIAL}
\end{center}
Supplementary Material discusses specific cases where traditional methods like MLE and MME become computationally infeasible due to limited sample sizes. It also includes comprehensive simulation results with complete tables and graphical analyses that demonstrate the performance of the fiducial approach covering both interval and point estimation, as well as detailed goodness‐of-fit assessments for the Rocke–Lorenzato calibration model on cadmium and copper datasets. The final section of the supplementary material includes the cadmium and copper datasets used in the manuscript.

\bibliographystyle{jabes}
\bibliography{references}

\end{document}